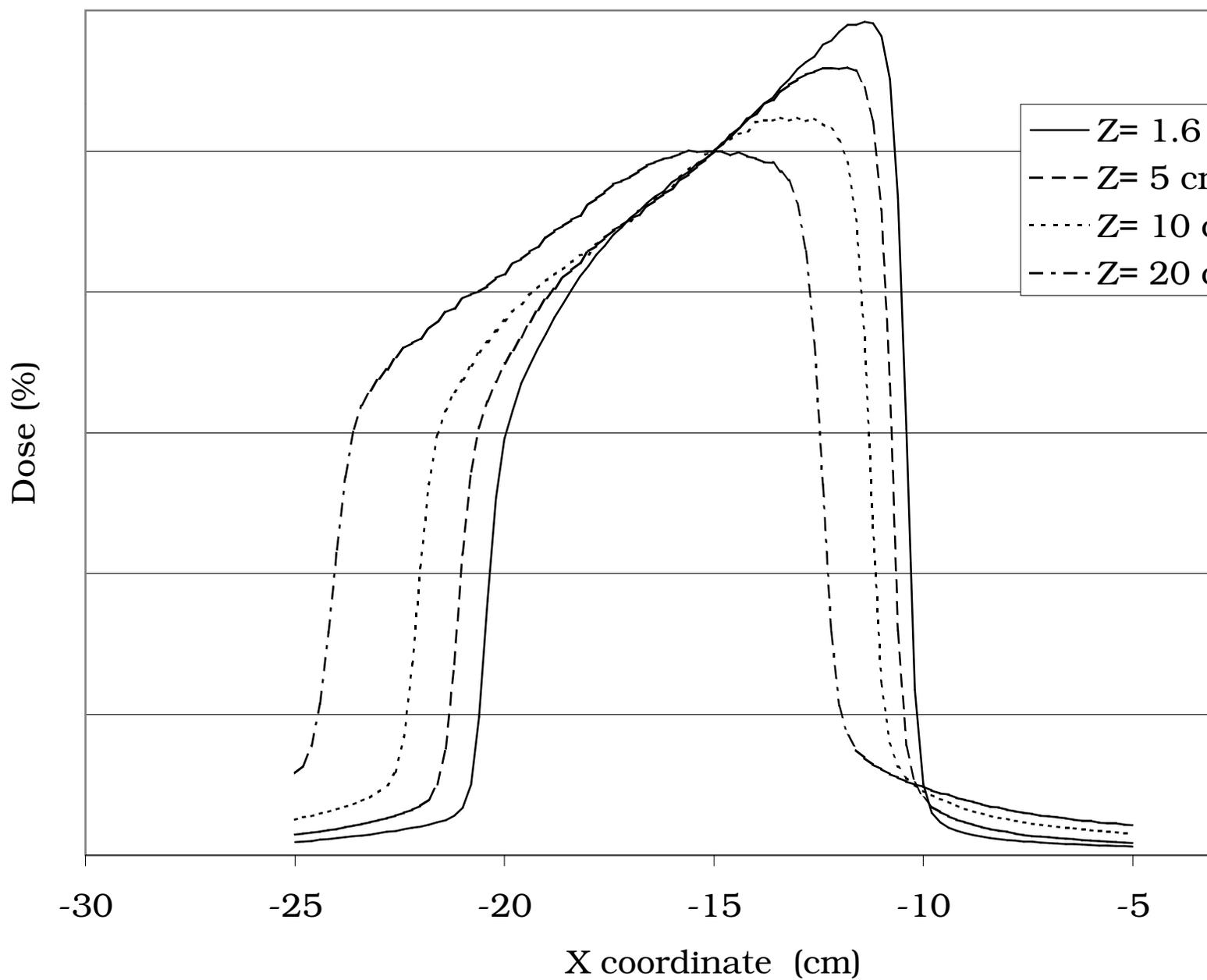

Figure 1(a). 6 MV X-profiles (Y= 15 cm.)

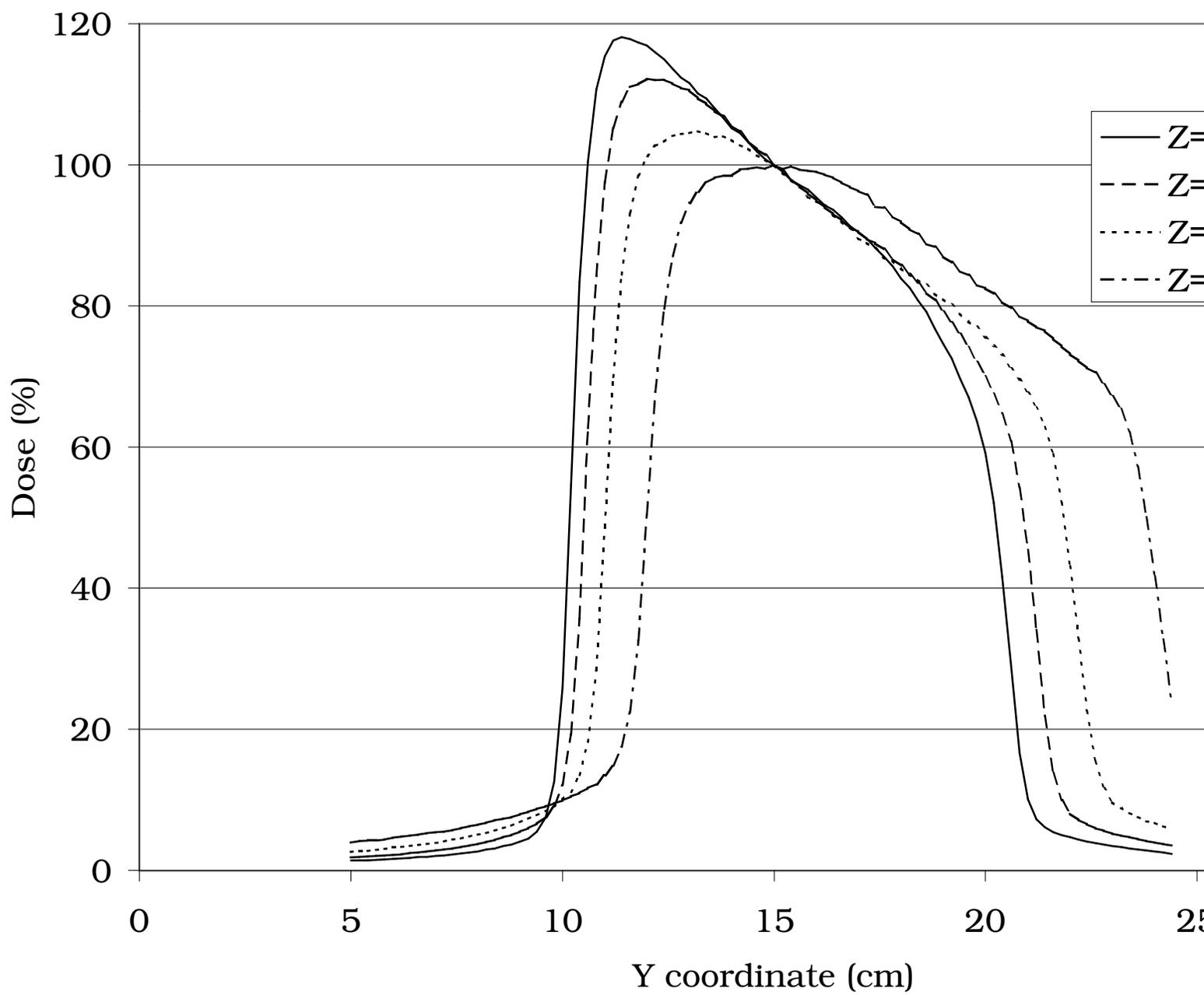

Figure 1(b). 6 MV Y-profiles (X= 15 cm.)

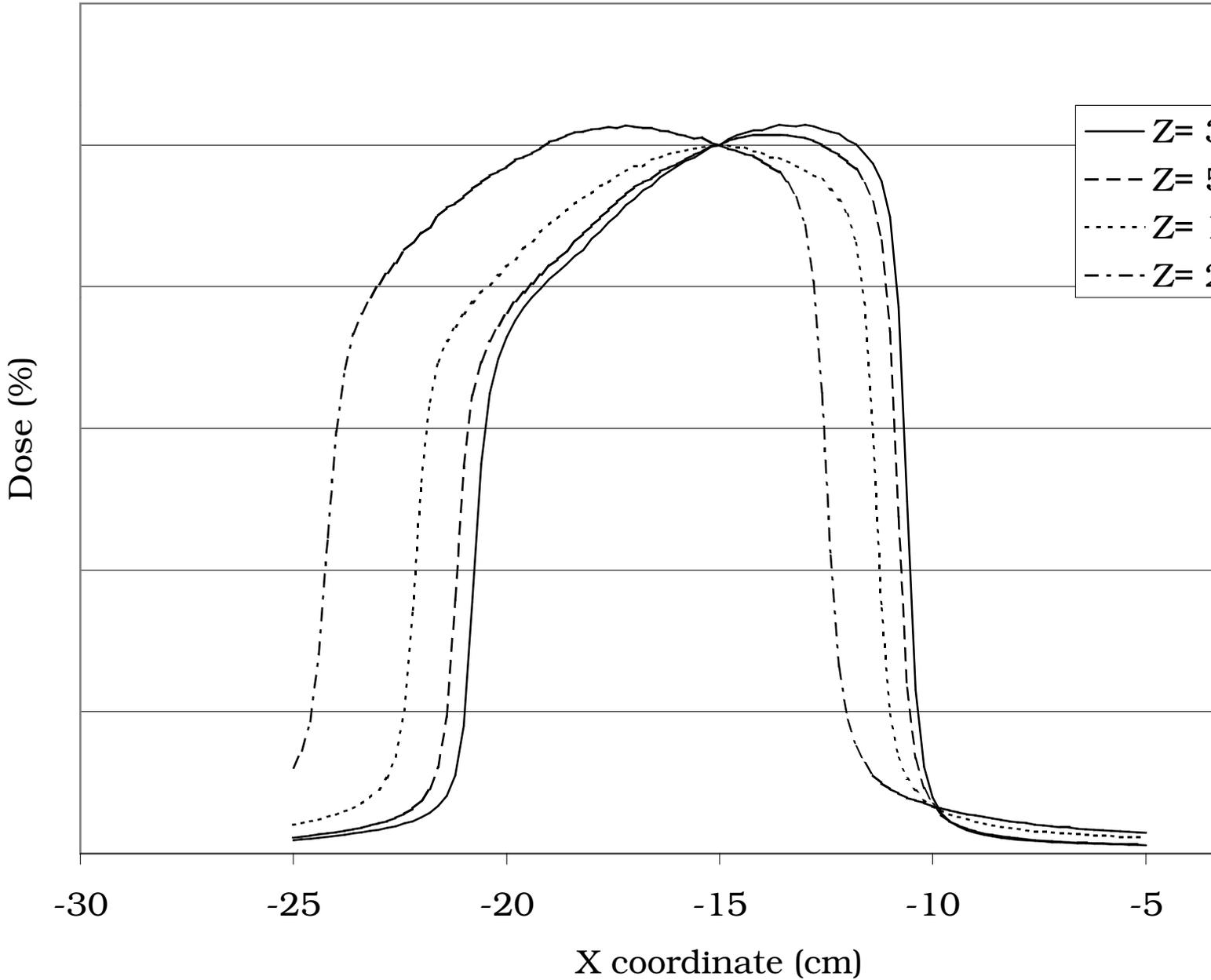

Figure 1(c). 18 MV X-profiles (Y= 15 cm.)

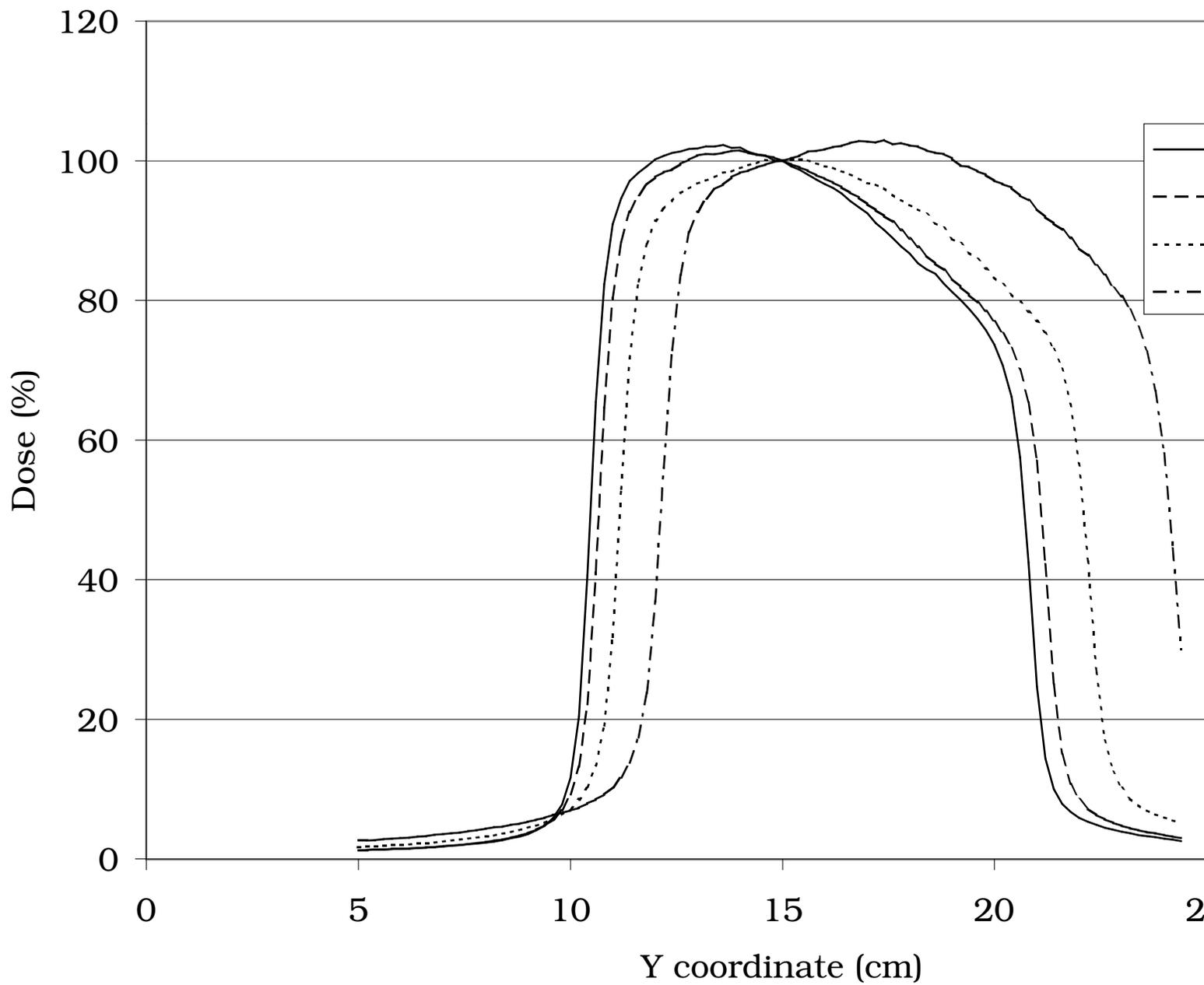

Figure 1(d). 18 MV Y-profiles (X= 15 cm.)

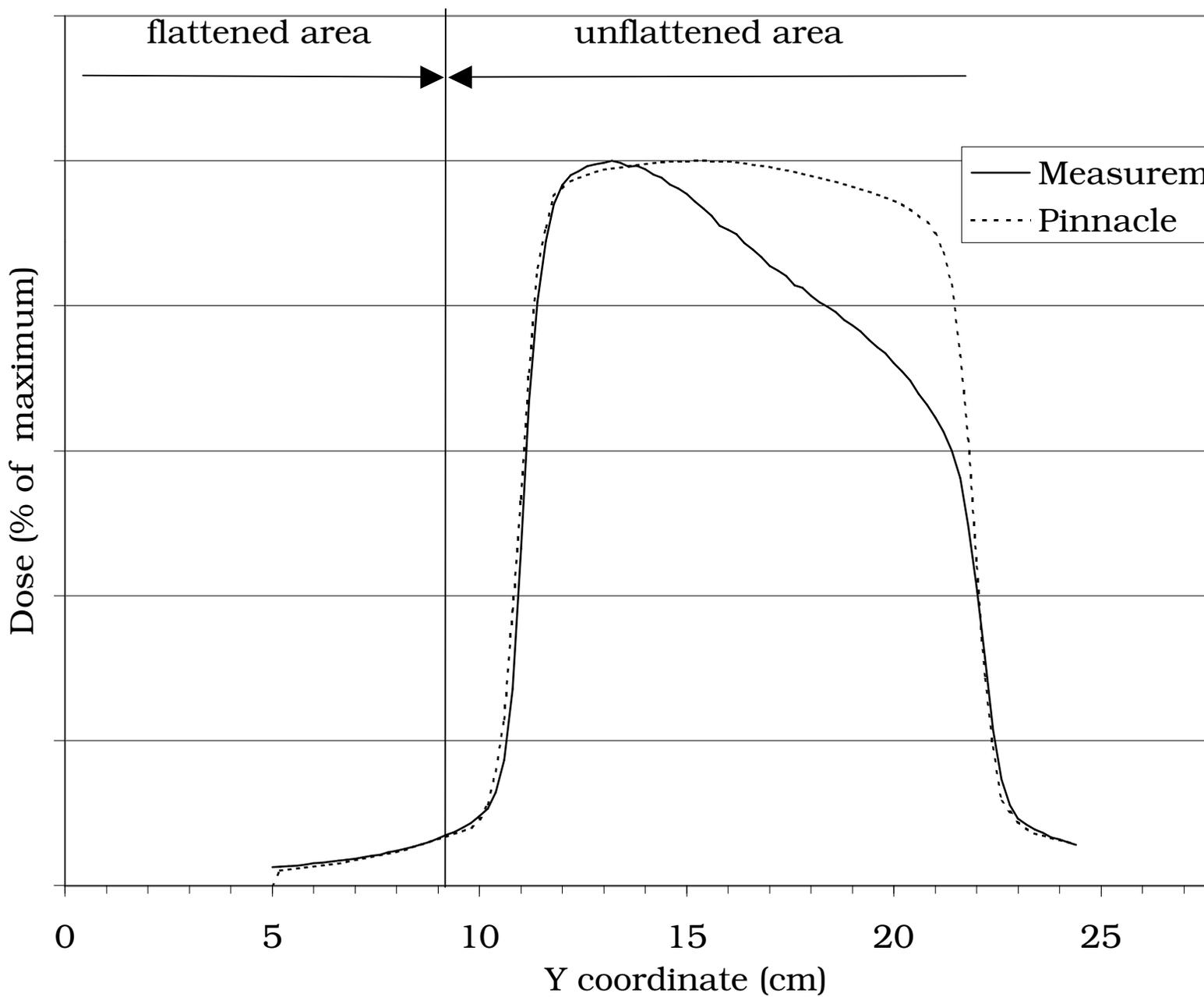

Figure 2(a). 6 MV Z = 10 cm. Measured and calculated

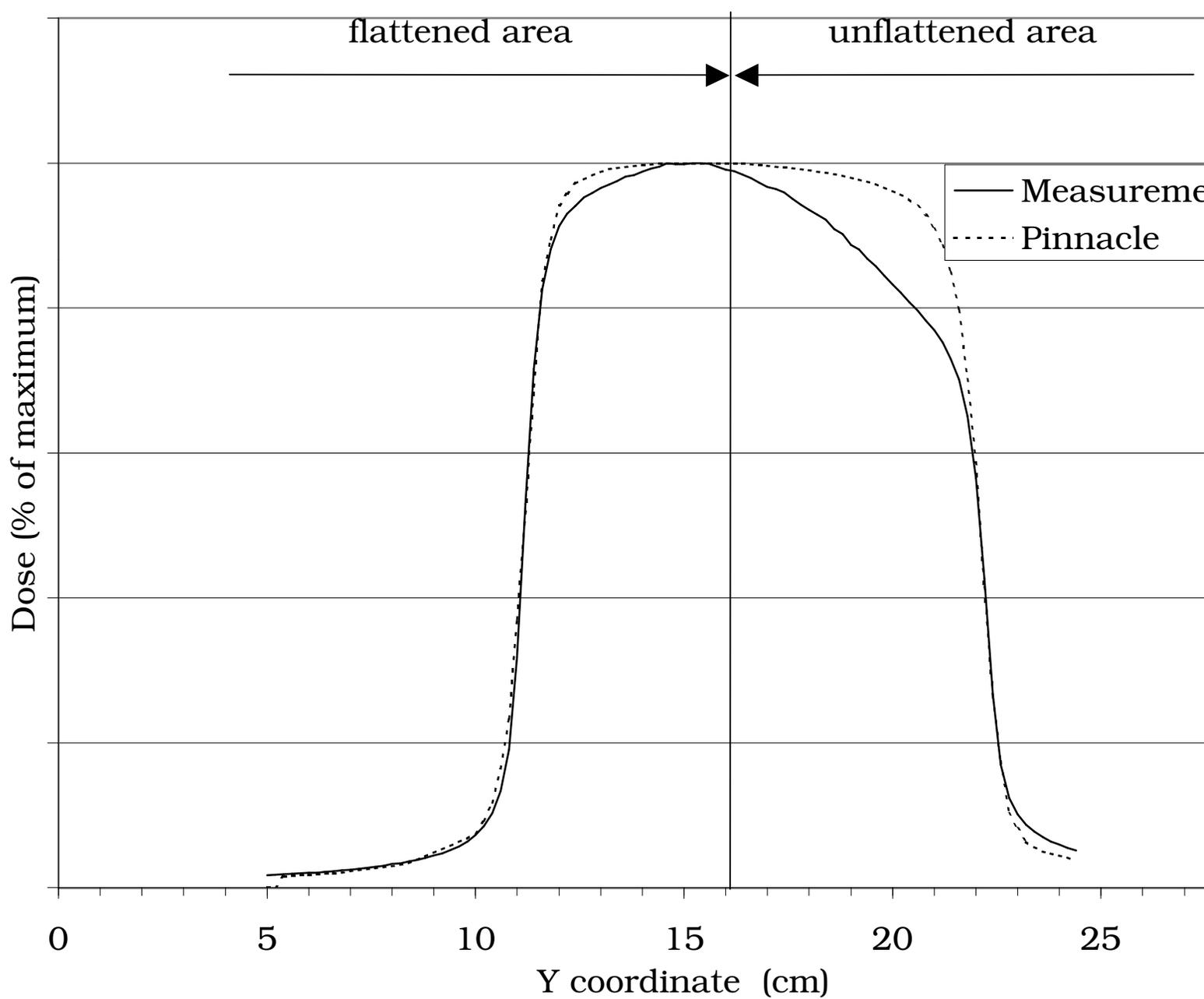

Figure 2(b). 18 MV Z = 10 cm Measured and calculated